\documentclass[sigconf, nonacm]{acmart}

\newcommand\vldbavailabilityurl{http://vldb.org/pvldb/format_vol14.html}

\begin{document}
\title{Zeoco: An insight into daily carbon footprint consumption}

\author{Karthik Ramakrishnan}
\affiliation{%
  \institution{Manipal Institute of Technology}
  \streetaddress{P.O. Box 1212}
  \city{Manipal}
  \state{India}
  \postcode{43017-6221}
}
\email{karthikramakrishnan14@gmail.com}

\author{Gokul P}
\affiliation{%
  \institution{Manipal Institute of Technology}
  \streetaddress{P.O. Box 1212}
  \city{Manipal}
  \state{India}
  \postcode{43017-6221}
}
\email{gokulpgns@outlook.com}

\author{Preet Batavia}
\affiliation{%
  \institution{Manipal Institute of Technology}
  \streetaddress{P.O. Box 1212}
  \city{Manipal}
  \state{India}
  \postcode{43017-6221}
}
\email{preetbatavia@gmail.com}

\author{Shreesh Tripathi}
\affiliation{%
  \institution{Manipal Institute of Technology}
  \streetaddress{P.O. Box 1212}
  \city{Manipal}
  \state{India}
  \postcode{43017-6221}
}
\email{tripathi.shreeshv@gmail.com}

\begin{abstract}
Climate change, which is now considered one of the biggest threats to humanity, is also the reason behind various other environmental concerns. Continued negligence might lead us to an irreparably damaged environment\cite{solomon2009irreversible}. After the partial failure of the Paris Agreement, it is quite evident that we as individuals need to come together to bring about a change on a large scale to have a significant impact. This paper discusses our approach towards obtaining a realistic measure of the carbon footprint\cite{wiedmann2008definition} index being consumed by a user through day-to-day activities performed via a smartphone app and offering incentives in weekly and monthly leader board rankings along with a reward system. The app helps ease out decision makings on tasks like travel, shopping, electricity consumption, and gain a different and rather numerical perspective over the daily choices.
\end{abstract}

\maketitle

\ifdefempty{\vldbavailabilityurl}{}{
\vspace{.3cm}
\begingroup\small\noindent\raggedright\textbf{Github Repository}\\
The source code, data, and/or other artefacts have been made available at \url{http://github.com/L3thal14/Zeoco}.
\endgroup
}

\section{Introduction}

Climate change \cite{solomon2009irreversible} is another human-made disaster responsible for killing more people than Covid-19 on the charts. Due to climate change, many annihilating disasters occur like melting glaciers, rising sea levels, and a myriad of more such unprecedented harmful effects of climate change. Despite commendable efforts by world leaders, multiple agreements have failed miserably. We all have known about climate change but are not working tenaciously to fight it. Humankind, while working in cognition, can do both good and bad to humankind as a whole. In retrospection, we can conclude that we are on the paragon of industrialization and automation and worked doggedly to achieve it. Still, on the same timeline, we have overlooked climate change and its deleterious effects. Humans have built astonishing things using technology to ease their everyday lives, but many of such inventions have proved to be destroying the environment such as automobiles. Thus it is of the prime importance to at least start at a personal level and contribute your best towards climate change using technology\cite{pandey2011carbon}, which can be the best tool at hand. 

We built an app called "Zeoco" that works towards helping every pedestrian reduce carbon emissions in the environment by checking on each product's carbon footprint and every product they purchase or use in their quotidian life. This app can prove to be a one-stop solution for everything and has been tailored very meticulously by observing the minutiae of a person's life and basic needs\cite{wackernagel1998our}. It has been made such that at every step, one can reduce carbon emissions. 
\\[0.5\baselineskip]
In contemporary times we have known that there is excessive use of vehicles and thus burning of a precious pool of fossil fuels to aid our travel. Although electric vehicles have started replacing a very small amount of petrol/diesel-run cars, it will take many more years to replace the fuel cars, which also seems quite quixotic ultimately. Our app provides a very efficient and exciting solution to this problem. Tracking your vehicle trip using GPS , whether you are in your private vehicle or taxi, based on the amount of distance travelled and taking information on the kind of car and fuel used , it calculates how much carbon has been emitted in the same and suggests smart alternatives such as walking, cycling, car pooling or using public transport based on your travel history. 
\\[0.5\baselineskip]
Food is the basic need of every living being. In modern times, we have seen a stark rise in packed and ready-made food, which can be very dangerous to the environment due to tin and other synthetic preservatives. Thus, our app provides a myriad of recipes to choose from with precalculated footprints and based on your search; it offers smarter and healthier alternatives that will help reduce the carbon footprint.
\\[0.5\baselineskip]
Use of barcode has been extensive in all the supermarkets and grocery stores to detect goods at the counter for payment. We have brought the barcode scanner to the app itself. Thus, the customer can scan the barcode of the product he/she is purchasing, and the app gives better alternatives having a lesser carbon footprint that the customer can choose. This also reflects on the screen of the cash counter, thus automating the whole process. 
\\[0.5\baselineskip]
We all tend to read through short news to imbibe more information about what is happening in and around. Thus, we have added real-time news based on climate change that the person can go through to instill a feeling of motivation \cite{kanemoto2016mapping}. 
\\[0.5\baselineskip]
Finally, we have added the most exciting part of the app, which is the leader board. We often tend to push ourselves to perform better where there is competition. Thus one can compete with their friends by truthfully using the app, and the one with the least carbon footprint tops the charts.

\begin{figure*}
  \includegraphics[width=5in]{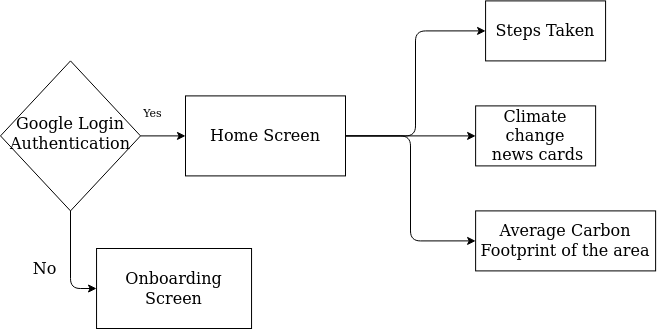}
  \caption{Authentication workflow}
\end{figure*}

\section{components}

Zeoco considers various factors to determine the total carbon footprint consumed by a person before comparing it with the average carbon footprint of the area and providing personalised tips. The final footprint values belonging to each of these factors is summed up together and saved in a Google Cloud Firebase database serving as a back-end for the app.  

\subsection{Restaurant Menu Recommendation}

Food consumption alone is responsible for a third of the carbon emissions in developed countries. The carbon footprint left behind varies from food item to food item\cite{virtanen2011carbon}. The restaurant’s data is collected from Zomato API and is fed into Zeoco’s shared database hosted on Firebase. The database includes the names of the food items in the menu, and we are using the Eaternity API, which uses MODEL techniques to break down the food items into the constituent ingredients. Using a well-researched look-up table from TEST, we calculate the amount of carbon in each food item. On the user’s side, when the user chooses a food item, we have used a recommendation system to recommend better food items in the same category but with a lesser carbon footprint. Directing people towards a more carbon-friendly menu will significantly reduce the negative environmental impact of hospitality.

\subsection{Travel Carbon Footprint Tracker}

Travel is an integral part of our daily lives. From traveling across continents via flights for holidays to traveling to work on motorbikes or cars, a massive amount of fuel is consumed\cite{holmatov2020environmental}\cite{kakouei2012estimation}\cite{li2019research}, emitting unwanted greenhouse gases, contributing to the user's daily carbon footprint consumption\cite{wei2017research}. Before every trip, the user needs to put in the destination, after which the app calls the Google Maps API in the back-end to calculate the total ride distance. Google provides a JavaScript Object Notation(JSON) file with credentials like the public and secret Maps API key, which is used to apply authentication and connect with the servers to send and receive data. This distance value in decimals is passed on to a form on the next page. The user needs to fill in the mode of transport and the type of fuel used. This information is used to call Trip to Carbon API, which calculates the carbon footprint based on available fuel usage and distance traveled information. The footprint value returned in the API's response is fed to the Firebase database with user permission adding to the overall carbon footprint consumed by the user.

\subsection{Electricity Consumption}

Electronic items ranging from toasters to laptops and televisions depend on a consistent electricity supply to run. If the global climate warms by one °C, the annual peak electricity consumption increases by 36.1\%, which goes on in a loop resulting in increased carbon consumption\cite{akhmat2014does}\cite{goldstein2020carbon}.
The user takes a picture of the monthly electricity billing document. We have deployed an optical character recognition(OCR)\cite{yindumathi2020analysis} model trained using a fair amount of datasets to the cloud, capturing and extracting the bill's total cost from the document. Geolocation API gets the location of the residence of the user. It is used to obtain the price per unit of electricity of that particular area and, therefore, the total electricity consumed in a month\cite{li2019climate}. This is passed on to calculate the carbon footprint consumed and added to the Firebase database. 

\begin{figure*}[ht] \setlength\fboxsep{0pt}
\hspace*{-1.5cm}%
\fcolorbox{white}{white}{\includegraphics[width=8.2in, height=8in]{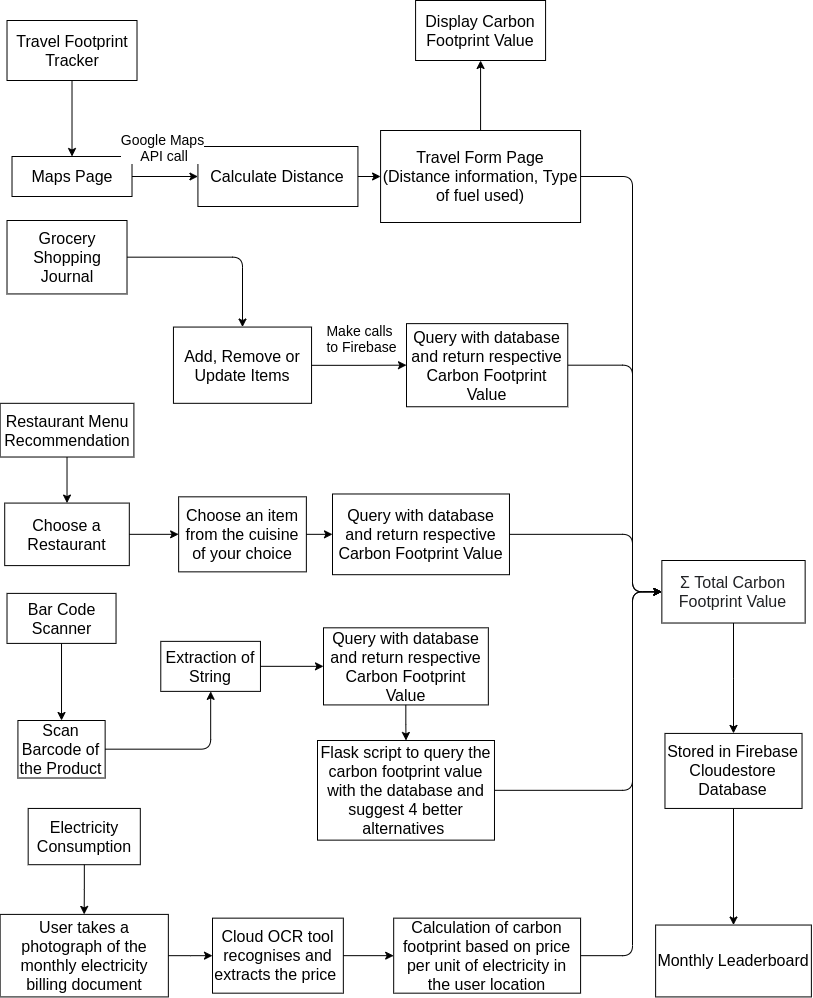}} \caption{A flow chart showing the whole workflow of Zeoco} \label{f7} \end{figure*}

\subsection{Bar Code Food Scanner}

One of the most tedious tasks to be fulfilled while shopping is the scanning of barcodes on the products\cite{pathak2010carbon}. Zeoco includes functionality wherein the user scans the barcode, and a unique string corresponding to the product is extracted and sent to the back-end built using Flask and stored in the Firebase database. This is then queried with a database containing the products available at the supermarket user is shopping in currently along with their carbon footprint values. Then, a script rearranges products depending on the food category the product belongs to and the carbon footprint value to provide four better alternatives to the user\cite{hartikainen2014finnish}. The user then has a choice to stick with the original choice or go ahead with one of the suggested alternatives\cite{hua2011managing}. This is passed on to calculate the carbon footprint consumed and added to the Firebase database.

\subsection{Grocery Shopping Items Journal}

This is a personal Shopping list Journal similar to a To-Do list but for Grocery shopping. The interface provides a simple provision to Create, Read, Update and Delete(CRUD) functionality to the user. As the user adds an item to the list just like adding an item to cart on an E-commerce store, in this case, instead of the price of the item, the corresponding carbon footprint is fetched and displayed. After an item is purchased, the user is presented with an option to add it to the total carbon footprint value present in the Firebase database.

\section{conclusions}

Science, accumulated and reviewed over decades, tells us that our planet is changing in ways that will have profound impacts on all of humankind. Now, we know climate change is only part of the problem, human ignorance being the greater contributor. The 12 warmest years in recorded history have all come in the last 15 years. The potential impacts go beyond rising sea levels. Farmers see crops wilted one year, washed away the next; and the higher food prices get passed on to you.
\\[0.5\baselineskip]
However, our obliviousness and ignorance has left this sad truth buried under petty front-page news headlines. The smoke of propaganda choking the ones who have been trying to voice the threats that mankind's careless actions pose to our present and coming future. 
\\[0.5\baselineskip]
Zeoco is a tiny bit to catalyze the big change with the ambitions to instil mindfulness within us and everyone around with regards to the extent to which the littlest of our actions affect our planet; the one which has been carrying us for a long long time without demanding anything in return.
\\[0.5\baselineskip]
Zeoco quantifies the implicit effect of consuming a product or utility with respect to the associated carbon footprint index - that serves as a metric - accumulated over various processes a product/utility undergoes to finally come into being to the end customer\cite{kause2019public}. Zeoco tries to be as interactive as possible to help you un-trivialize decision-making from tasks like travel, shopping, electricity consumption and help you gain a different and rather a numerical perspective over your daily choices.

\section{Future Work}
We plan to partner with multiple restaurants and stores in order to increase our database size so that people have access to a wide variety of items. Future studies could involve a detailed study of carbon footprint consumption by users staying in different locations like metropolitan cities, countryside and small towns with respect to various factors captured by our app. Enhancing the User Interface(UI) and User Experience(UX) to ensure a minimum number of clicks to add the carbon footprint values smoothly is definitely in the pipeline. With the increase in user base, moving to cloud platforms with better request handling like Amazon EC2, Digital Ocean etc. to deploy our back-end models and scripts is an important step to be taken. Furthermore, for integration with the app, QR code cards can be installed or distributed at supermarkets and through cab sharing apps by getting access to their Application Programming Interface(API), making it easier for users. 

\bibliography{zeoco}

\bibliographystyle{icml2020}

\end{document}